\documentclass[conference, a4paper,10pt]{IEEEtran}

\IEEEoverridecommandlockouts
\usepackage{cite}
\usepackage{amsmath,amssymb,amsfonts}
\usepackage{algorithmic}
\usepackage{graphicx}
\usepackage{textcomp}
\usepackage[dvipsnames]{xcolor}
\def\BibTeX{{\rm B\kern-.05em{\sc i\kern-.025em b}\kern-.08em
    T\kern-.1667em\lower.7ex\hbox{E}\kern-.125emX}}

\usepackage{graphicx}
\usepackage{enumitem}

\usepackage{glossaries}
\newacronym{qos}{QoS}{quality of service}
\newacronym{urllc}{URLLC}{ultra-reliable low-latency communications}
\newacronym{rx}{RX}{receiver}
\newacronym{tx}{TX}{transmitter}
\newacronym{csi}{CSI}{channel state information}
\newacronym{5g}{5G}{5th generation mobile network}
\newacronym{ar}{AR}{augmented reality}
\newacronym{gp}{GP}{Gaussian process}
\newacronym{snr}{SNR}{signal-to-noise ratio}
\newacronym{cdi}{CDI}{channel distribution information}
\newacronym{bs}{BS}{base station}
\newacronym{ue}{UE}{user equipment}
\newacronym{cdf}{CDF}{cumulative distribution function}
\newacronym{ldg}{LDG}{Linear Congruential Generator}
\newacronym{vna}{VNA}{vector network analyzer}
\newacronym{cfr}{CFR}{channel frequency response}
\newacronym{if}{IF}{intermediate frequency}
\newacronym{cir}{CIR}{channel impulse response}

\newcommand{\mbf}[1]{\mathbf{#1}}
\newcommand{\mbs}[1]{\boldsymbol{#1}}
\newcommand{\mR}{\mathbb{R}}
\newcommand*{\cond}{\hspace*{1pt} |\hspace*{1pt}}

\newcommand{\mC}{\mathbb{C}}

\begin{document}

\title{Experimental Study of Spatial Statistics for Ultra-Reliable Communications
}

\author{%
\IEEEauthorblockN{Tobias~Kallehauge, Anders~E.~Kalør, Fengchun~Zhang, and Petar~Popovski}%
\IEEEauthorblockA{Department of Electronic Systems, Aalborg University, Denmark
(\{tkal,aek,fz,petarp\}@es.aau.dk)}}

\maketitle

\begin{abstract}
  This paper presents an experimental validation for prediction of rare fading events using channel distribution information (CDI) maps that predict channel statistics from measurements acquired at surrounding locations using spatial interpolation. Using experimental channel measurements from 127 locations, we demonstrate the use case of providing statistical guarantees for rate selection in ultra-reliable low-latency communication (URLLC) using CDI maps. By using only the user location and the estimated map, we are able to meet the desired outage probability with a probability between $93.6$--$95.6\%$ targeting $95\%$. On the other hand, a model-based baseline scheme that assumes Rayleigh fading meets the target outage requirement with a probability of $77.2\%$. The results demonstrate the practical relevance of CDI maps for resource allocation in URLLC.
\end{abstract}


\section{Introduction}
Future wireless systems are expected to deliver tailored connectivity services that provide ultra-high reliability availability and low latency~\cite {giordani20206G}. However, meeting these requirements while providing satisfiable data rate is a significant challenge due to the random nature of the wireless channel, which depends on the specific environment in which the system is deployed. A particularly challenging scenario arises when the instantaneous channel through \gls{csi} is unavailable, e.g., due to low latency requirements, sporadic transmissions, or high-mobile scenarios, where the \gls{csi} can quickly become outdated \cite{Chen2022CSI}. An alternative to relying on \gls{csi} is to choose the communication parameters according to statistical
knowledge of the channel, i.e., \gls{cdi}, so that requirements are met probabilistically with given confidence \cite{Angjelichinoski2019statistical}. The available resources for \gls{cdi} estimation are often limited, which may lead to failure in fulfilling the requirements or extreme overprovisioning of the system to account for the lack of information~\cite{Angjelichinoski2019statistical}, so a promising direction is to exploit spatial correlation of channel statistics using \emph{\gls{cdi} radio maps}~\cite{Kulzer2021CDI} (or simply \gls{cdi} maps) to estimate \gls{cdi} only based on location. 

A \gls{cdi} map is a spatial model of the channel statistics that are relevant for the given application, and can be created by spatially interpolating measurements taken across a region of interest. Once the map has been constructed, the channel statistics can be predicted at any location without additional measurements (other than location information). Enabling such maps is becoming increasingly feasible with the popularity of localization and sensing, and similar concepts have been defined in recent years, including \textit{statistical radio maps} \cite{kallehauge2022Globecom} and channel knowledge maps \cite{Zeng2021knowledge}. In contrast to classical radio maps (or coverage maps), which model average, large-scale parameters such as pathloss and shadowing~\cite{Chowdappa2018}, \gls{cdi} maps model higher-order statistics, allowing them to capture, e.g., small-scale fading or interference dynamics.

The use of \gls{cdi} maps and similar concepts for predictive resource allocation and ensuring communication requirements is already seen in the literature \cite{Kulzer2021CDI,Zeng2021knowledge}, and especially in the area of \gls{urllc} \cite{Azari2019Risk,kallehauge2022Globecom,Kallehauge2023deliver}. However, these works are primarily theoretical or 
based on simulated data, and there is a large gap in
experimental research on \gls{cdi} maps for predicting small-scale fading behavior. The few existing works tend to assume either Rayleigh or Rician fading~\cite{Yu2020river, He2023Train}, which may not accurately capture the tail probabilities needed for \gls{urllc} \cite{Angjelichinoski2019statistical}. Additionally, measuring small-scale fading is time-consuming, as it requires moving the antenna to many locations around a central point to emulate small-scale fading. As a result, these works tend to have only around 100 fading measurements per location, which is insufficient to capture the rare events needed for \gls{urllc} \cite{He2023Train,Rappaport2017smallscale}. 

In this paper, we provide an experimental validation of the use of \gls{cdi} maps to predict the small-scale fading statistics relevant to \gls{urllc}. Furthermore, we demonstrate that for a \gls{urllc} rate selection task, the use of a \gls{cdi} map significantly increases the probability that a predefined target outage probability is met. Finally, to acquire a sufficient number of samples, we utilize a wideband channel sounder, exploiting the duality between changes in frequency and small perturbations in space. This allows us to collect $8001$ narrowband fading measurements per channel sounding.

The remainder of the paper is structured as follows. Section \ref{sec:theory} presents the system model and problem statement. The method for the rate selection is given in Sec. \ref{sec:rate_selection}, and Sec. \ref{sec:experimental} details the theory and our experimental setup for measuring small-scale fading. Section \ref{sec:results} presents the collected data and results for rate selection in different scenarios, and finally, the paper is concluded in Sec. \ref{sec:conclusion}. 

\section{System Model} \label{sec:theory}
\subsection{Signal and fading model} \label{subsec:model}
We consider a single \gls{bs} covering a region $\mathcal{R}\subset\mR^3$. The \gls{bs} is located at a fixed location $\mbf{x}_{\text{BS}} \in \mathcal{R}$, and serves \glspl{ue} with low latency and ultra-high reliability requirements, which prevents estimation of the instantaneous \gls{csi}. Both the \glspl{ue} and the \gls{bs} are equipped with single antennas communicating over narrowband block-fading channels, which represent a worst-case scenario of small-scale fading due to low spatial and frequency diversity. 

For simplicity of exposition, we limit our focus to communication in the uplink, but our methodology applies to the downlink as well. Specifically, a normalized packet $\mathbf{s} \in \mC^n$ ($E[(1/n)\|\mathbf{s}\|_2^2] = 1$) transmitted with energy $P_{\text{tx}}$ by a \gls{ue} located at location $\mbf{x} \in \mathcal{R}$ is received by the \gls{bs} as 
\begin{align}
    y = \sqrt{P_{\text{tx}}}h(\mbf{x})\mathbf{s} + \mathbf{z}, \label{eq:channel}
\end{align}
where $h(\mbf{x}) \in \mC$ is the complex channel coefficient, which is assumed to be random and independently drawn from an unknown distribution that depends on the \gls{ue}'s location $\mbf{x}$, and $\mathbf{z}\overset{\text{iid}}{\sim}\mathcal{CN}(0,BN_0)$ is additive white Gaussian noise assuming a bandwidth $B$ and noise power spectral density $N_0$. The instantaneous \gls{snr} of the channel is given by $\gamma(\mbf{x}) = |h(\mbf{x})|^2\gamma_{\text{tx}}$, where $\gamma_{\text{tx}} = P_{\text{tx}}/(BN_0)$ is the transmit \gls{snr}. 
Since the \gls{ue} does not have \gls{csi}, it transmits with a fixed rate denoted by $R$ [bits/s/Hz]. We assume that packet errors occur due to outage, which happens when the instantaneous rate supported by the channel falls below $R$, i.e.,
\begin{align}
    p_{\text{out},\mbf{x}}(R) = P(\log_2(1 + \gamma(\mbf{x})) \leq R) = F_{\rho(\mbf{x})}\left(\frac{2^R - 1}{\gamma_{\text{tx}}}\right), \label{eq:p_out}
\end{align}
where $F_{\rho(\mbf{x})}$ is the \gls{cdf} of the channel gain $\rho(\mbf{x}) = |h(\mbf{x})|^2$. Note that the packet error probability in \eqref{eq:p_out} is accurate even in the short blocklength regime~\cite{yang14quasistatic}. It is assumed for simplicity that the fading process characterized by $F_{\rho(\mbf{x})}$ is time stationary --- see, e.g., \cite{Zeng2021knowledge} for discussion on \gls{cdi}-maps in non-stationary environments.

\subsection{Problem statement} \label{subsec:problem}

We next consider the problem of selecting the maximum rate $R$, that guarantees that the outage probability is at most $\epsilon \in (0,1)$ for a \gls{ue} located at an arbitrary but known location $\mathbf{x}$. When the distribution of $\rho(\mbf{x})$ is perfectly known, the optimal rate is given by the \textit{$\epsilon$-outage capacity}, defined as
\begin{align}
    R_{\epsilon}(\mbf{x}) = \sup \{R \cond p_{\text{out},\mbf{x}}(R) \leq \epsilon\}= \log_2(1+ F_{\rho(\mbf{x})}^{-1}(\epsilon)), \label{eq:out_capacity}
\end{align}
where $F_{\rho(\mbf{x})}^{-1}(\epsilon)$ is the $\epsilon$-quantile of $\rho(\mbf{x})$. However, when the fading distribution is estimated using a finite number of samples, this definition is not meaningful due to the inherent randomness in the dataset. 

To model the process of learning the fading distribution from spatial samples, we examine a scenario where the \gls{bs} has collected a dataset $\mathcal{D}=\{\mbs{\rho}_d,\mbf{x}_d\}_{d=1}^D$ comprising for each location $\mbf{x}_d\in\mathcal{R}$, $d = 1,\dots, D$, a set of $N$ independent small-scale fading channel measurements $\mbs{\rho}_d=  \left( \rho_{d,1} \ \dots \ \rho_{d,N} \right)$ (i.e., a total of $ND$ measurements). A \gls{ue} at a previously unobserved location $\mbf{x} \in \mathcal{R}\setminus \{\mbf{x}_d\}_{d=1}^D$  now joins the network and needs to select its communication rate based on the data, denoted  $R(\mbf{x} \cond \mathcal{D})$. This scenario includes several sources of uncertainty, including the underlying propagation environment in the cell characterized by $F_{\rho(\mbf{x})}$, the data $\mathcal{D}$, and the \gls{ue} location $\mbf{x}$. By modeling the uncertainties as random processes, we employ the statistical learning approach of characterizing reliability through the \textit{meta-probability} \cite{kallehauge2022Globecom}, i.e., the probability of exceeding the reliability requirement: 
\begin{align}
    \tilde{p}_{\epsilon} = P\left(p_{\text{out},\mbf{x}}(R(\mbf{x} \cond \mathcal{D})) > \epsilon \right),  \label{eq:MetaConstraint}
\end{align}
where the outer probability accounts for \textit{all} uncertainties in the scenario, which includes both uncertainty in the spatial domain and in the random channel measurement at each location. Note that location uncertainty has been omitted here for simplicity; hence $\mbf{x}$ is considered the true location --- see \cite{kallehauge2023statistical} for analysis on the impact of localization error on reliability. Defining $\delta > 0$ as the \textit{confidence parameter}, we pose \emph{the problem of finding a location-based rate selection approach that satisfies}
\begin{align}
 \tilde{p}_{\epsilon}(\mathbf{x}) \leq \delta, \label{eq:problem}
\end{align}
hence bounding the probability of exceeding the reliability requirement. Fulfilling \eqref{eq:problem} has some trivial solutions, e.g., $R(\mbf{x} \cond \mathcal{D}) = 0$, so  we evaluate the performance of the rate selection function through the \textit{normalized throughput} \cite{kallehauge2022Globecom}: 
\begin{align}
    \tilde{R}_{\epsilon}(\mbf{x}) =  \frac{R(\mbf{x}\cond \mathcal{D})(1-p_{\text{out},\mbf{x}}(R(\mbf{x}\cond \mathcal{D}))}{R_{\epsilon}(\mbf{x})(1-\epsilon)}, \label{eq:throughput}
\end{align}
which is exactly $1$ if the predicted outage capacity is perfectly estimated and, hence, $R(\mbf{x}\cond \mathcal{D}) = R_{\epsilon}(\mbf{x})$. 

\section{Rate Selection using Channel Distribution Information Maps}\label{sec:rate_selection}
In this section, we outline the considered rate selection method proposed in~\cite{kallehauge2022Globecom}, which relies on the construction of a \gls{cdi} map. This method has the advantage that it does not require a specific parametric distribution, such as Rician or Rayleigh, but instead directly models the $\epsilon$-quantile of the fading distribution as a Gaussian process. The procedure is divided into three steps presented in the sequel.

The first step is to estimate the log-fading power quantile at each location in $\mathcal{D}$ as \cite{Ord1994} 
\begin{align}
    \widehat{q}_{\epsilon,d} = \ln\left({\rho}_{d,(r)}\right), \quad r = \lfloor N\epsilon \rfloor,\quad d = 1,\dots,D \label{eq:quantile_est}
\end{align}
where $\rho_{d,(r)}$ is the $r$-th order statistics of $\mbs{\rho}_d$ and $\lfloor \cdot \rfloor$ is the floor function. 

The next step is to construct the map via spatial prediction/interpolation, i.e., estimating $q_{\epsilon}(\mbf{x})$ for all $\mbf{x}$ in the cell by modeling the quantiles as a Gaussian process. We employ the framework from our previous work \cite{kallehauge2022Globecom}, which allows us to obtain a \textit{predictive distribution} of the quantiles, which is central for the statistical learning framework. For the sake of brevity, we omit the details of the prediction framework, and we refer to \cite[Sec. III-B]{kallehauge2022Globecom} for a complete description. In short, the Gaussian process framework takes $\mathcal{D}$ and $\mbf{x}$ as inputs, estimates various hyperparameters, and finally provides a predictive distribution of the quantile at location $\mbf{x}$. The predictive distribution is given by a Gaussian distribution
\begin{align}
    q_{\epsilon}(\mbf{x} \cond \mathcal{D}) \sim \mathcal{N}(\mu(\mbf{x} \cond \mathcal{D}), \sigma^2(\mbf{x} \cond \mathcal{D})), \label{eq:pred_dist}
\end{align}
where $\mu$ is the \textit{predictive mean} and $\sigma^2$ is the \textit{predictive variance}, characterizing the uncertainty of the prediction at location $\mbf{x}$ based on $\mathcal{D}$. 

The final step is to select the communication rate based on the predictive distribution. We start by re-writing the meta-probability \eqref{eq:MetaConstraint} using \eqref{eq:p_out}: 
\begin{align}
    \tilde{p}_{\epsilon} &= P\left(F_{\rho(\mbf{x})}\left(\frac{2^{R(\mbf{x} \cond \mathcal{D})} - 1}{\gamma_{\text{tx}}}\right) > \epsilon \right) \nonumber \\
    &= P\left(\frac{2^{R(\mbf{x} \cond \mathcal{D})} - 1}{\gamma_{\text{tx}}}> F_{\rho(\mbf{x})}^{-1}(\epsilon) \right) \nonumber \\
    &= P(R(\mbf{x} \cond \mathcal{D}) > \log_2(1+\gamma_{\text{tx}}e^{q_{\epsilon}(\mbf{x})})), \label{eq:meta_deriv}
\end{align}
using the identity $F_{\rho(\mbf{x})}^{-1}(\epsilon) = e^{F_{\ln\rho(\mbf{x})}^{-1}(\epsilon)} = e^{q_{\epsilon}(\mbf{x})}$ in the last equation. The rate is then selected as the $\delta$-quantile of the predictive distribution of $q_{\epsilon}$:
\begin{align}
    &R(\mbf{x} \cond \mathcal{D}) = \log_2\left(1 + \exp\left(F_{q_{\epsilon}(\mbf{x} \cond \mathcal{D})}^{-1}(\delta)\right)\right) \label{eq:rate_select}  \\
    &= \log_2\left(1 + \exp\left(\mu(\mbf{x} \cond \mathcal{D}) + \sqrt{2}\sigma(\mbf{x} \cond \mathcal{D})\mathrm{erf}^{-1}(2\delta - 1)\right)\right),  \nonumber
\end{align}
where $\mathrm{erf}^{-1}$ is the inverse error function. Inserting \eqref{eq:rate_select} into \eqref{eq:meta_deriv} yields
\begin{align}
    \tilde{p}_{\epsilon} &= P(F_{q_{\epsilon}(\mbf{x} \cond \mathcal{D})}^{-1}(\delta) > q_{\epsilon}(\mbf{x})) = F_{q_{\epsilon}(\mbf{x})}(F_{q_{\epsilon}(\mbf{x} \cond \mathcal{D})}^{-1}(\delta)) \approx \delta,
\end{align}
where the last equality holds under the assumption that the quantile follows a Gaussian process. Although this is a strong assumption, the experimental results in Sec. \ref{sec:results} demonstrate that it generally performs well.

\section{Experimental Measuring of Small-Scale Fading} \label{sec:experimental}

The problem stated in Sec. \ref{subsec:problem} assumes the availability of a dataset $\mathcal{D}$ comprising several independent small-scale fading measurements for different locations in an area. This section first provides a theoretical background for measuring small-scale fading and then describes the experimental setup used in our measurement campaign. 
\subsection{Frequency-domain sampling of small-scale fading} \label{subsec:small_scale}
Small-scale fading is a result of random variations in the phases of the individual signal paths between the transmitter and the receiver. The resulting narrowband channel response can be modeled as a tapped-delay line, assuming that the signal arrives at the destination from a fixed number of (unresolvable) paths with different delays. Let $K$ denote the number of paths, and let $\alpha_k \in \mC$, $\tau_k > 0$ denote the coefficient and delay for each path $k = 1,\dots, K$, respectively. The tapped-delay line model for narrowband transmission at frequency $f$ is then \cite{Goldsmith2005}
\begin{align}
    h = \sum_{k=1}^K \alpha_ke^{-2\pi j f \tau_k} = \sum_{k=1}^K \alpha_ke^{-2\pi j d_k/ \lambda },  \label{eq:tapped_delay}
\end{align}
where $d_k = \tau_k\cdot c$ is the distance traveled along path $k$, and $\lambda = c/f$ is the wavelength (the dependence on the device location $\mbf{x}$ is omitted for brevity). It is seen that small changes in location (relative to the wavelength) can cause constructive and destructive interference between the different paths, which can significantly change the channel gain and, hence, induce the small-scale fading effect. The key assumption here is that location changes only affect delays of the individual paths while the magnitudes of the coefficients remain roughly constant, i.e., $\alpha_k(\mbf{x} + \Delta \mbf{x}) \approx \alpha_k$ for small perturbations $\Delta\mbf{x}$. 

Similarly, the model \eqref{eq:tapped_delay} shows that changing frequency also affects the phases for the different paths. Assuming that the coefficients remain roughly constant across frequency, i.e., $\alpha_k(f + \Delta f) \approx \alpha_k$ for frequency changes $\Delta f$ and that the overall statistical process is stationary, collecting samples in random frequency bands is thus roughly equivalent to collecting samples over random locations. 
To realize a wide range of fading samples with the frequency approach, it is required that the range of different frequencies is large compared to the coherence bandwidth of the channel\footnote{Similarly, location-based measurements require a large range of distances compared to the coherence distance of the channel.}. Hence, with frequency range $[f_{\min}, f_{\max}]$ and coherence bandwidth $B_c$, then $f_{\max} - f_{\min} \gg B_c$ is sufficient to simulate small-scale fading by drawing $f$ uniformly in the frequency range. 

Based on the discussion above, we conclude that the small-scale fading distribution can be approximated by samples obtained in the frequency domain. However, compared to location-based sampling, which is time-consuming due to mechanical movements, sampling the frequency domain using wideband channel sounding can be performed in short time. As a result, more samples can be obtained, which in turn enables a better approximation of the fading distribution.

\subsection{Experimental setup}

The measurements were conducted in a courtyard enclosed by buildings, giving a rich scattering environment with the \gls{rx} placed to give a combination of line-of-sight and non-line-of-sight channels. 
The measurement system comprised the following components: 
\begin{itemize}[leftmargin=*]
	\item A \gls{tx} antenna: Homemade bi-conical antenna as the \gls{tx} antenna, detailed in \cite{zhekov2016modified}. This antenna features vertical polarization and covers the frequency range from $1.5$ to $41$ GHz with a gain within the range of $1.2$ to $5.1$ dB. 
	\item A \gls{rx} antenna: Bi-conical antenna (INFO SZ-2003000/P) with vertical polarization. It covers the frequency range from $2$ to $30$ GHz, with a gain within the range of $1$ to $7$ dB. 
	\item A \gls{vna}: An Anritsu ShockLine MS46131A-043 VNA was employed to measure the \gls{cfr} between the \gls{tx} and \gls{rx} antenna ports. The VNA was configured with operating frequencies ranging from $2$ to $10$ GHz, utilizing $8001$ frequency points, a transmitted power level of $0$ dBm, and an \gls{if} bandwidth of $1000$ Hz.
	\item A laptop: To control the operation of the VNA and store the measurement data.
\end{itemize}

\begin{figure}
    \centering
    \includegraphics[width = \linewidth]{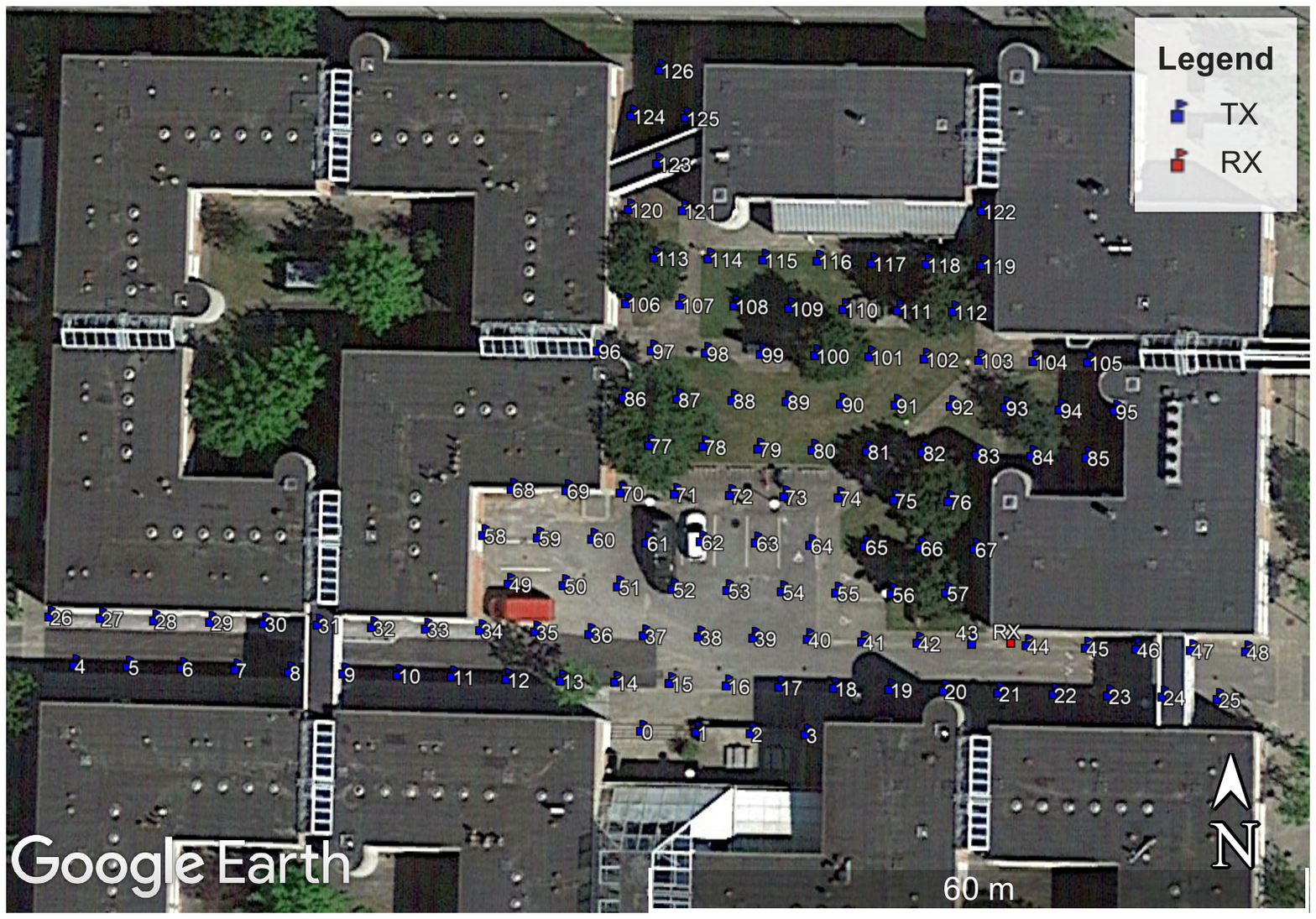}
	\caption{\Gls{rx} location (in red) and 127 different \gls{tx} locations (in blue).}
	\label{fig:map}
 \vspace{-4mm}
\end{figure}

During the measurement process, the \gls{rx} antenna remained fixed, while the \gls{tx} antenna was positioned at $127$ different locations, forming a uniform grid of equilateral triangles with sidelength $5$ m as illustrated in Fig. \ref{fig:map}. For each \gls{tx} location, the \gls{vna} recorded the \gls{cfr} between the \gls{tx} and the \gls{rx} antenna ports. The power of the frequency response at each frequency also approximates the power of a narrowband channel at that particular frequency, hence giving $8001$ fading measurements per location under the assumptions in the previous section.  

\section{Experimental results} \label{sec:results}
In this section, we present the results of estimating \gls{cdi} maps and performing predictive rate selection using the data from the measurement campaign. We start by validating the experimental data.

\subsection{Data validation}  \label{subsec:data}
\begin{figure*}
	\centering
 \begin{minipage}{0.38\textwidth}
    \centering
    \vspace{-3mm}
    \includegraphics[scale = 0.5]{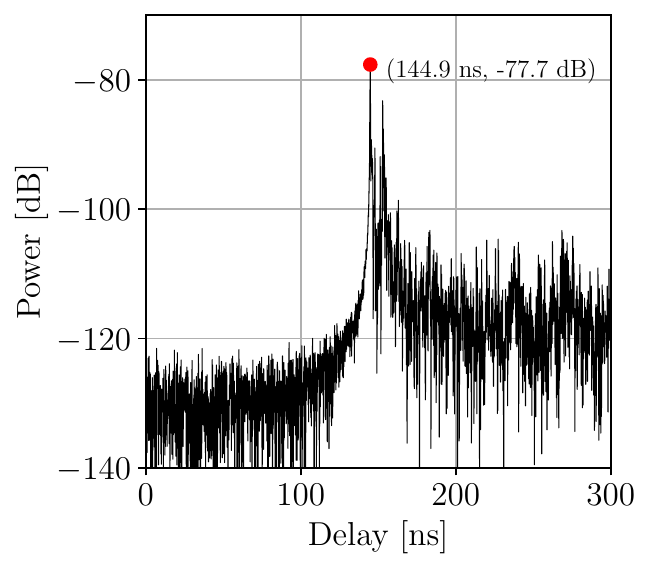}
	\caption{CIR at the \gls{tx} location 35.}
	\label{fig:cir_0}
 \end{minipage}
\begin{minipage}{0.6\textwidth}
    \centering
    \includegraphics[scale = 0.5]{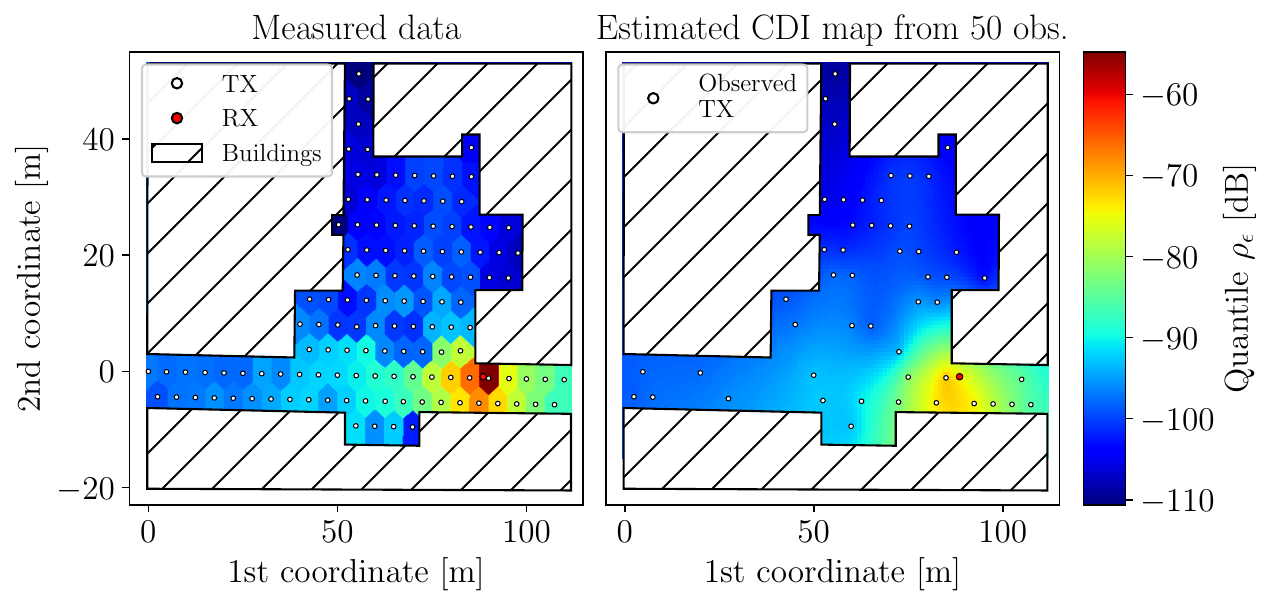}
    \caption{$\epsilon$-quantile of fading $\rho_{\epsilon} = F_{\rho}^{-1}(\epsilon)$ with $\epsilon = 1\%$ in dB. The left plot shows the measured data at the \gls{tx} locations interpolated based on the closest measured point. The right plot shows values predicted by a high-resolution \gls{cdi} map based on $50$ observations.}
    \label{fig:quantile_map}
\end{minipage} 
\end{figure*}

To validate the reliability of the measurement data, we performed an inverse Fourier transformation on the \gls{cfr} recorded at the \gls{tx} location 35 to obtain the corresponding \gls{cir}, depicted in Fig. \ref{fig:cir_0}. The CIR reveals that the flight time of the direct path between the \gls{tx} and the \gls{rx} is $144.9$ ns, corresponding to a distance of $43.4$ m. This distance aligns well with the measured distance of $43.6$ m. We can further validate the data by comparing the received power of the direct path to the theoretical free-space path loss. At the center frequency of 6 GHz, the free-space power loss is given as $20\log_{10} \frac{c}{4\pi d f_c }$, where $c$, $d$, and $f_c$ is the speed of light, the flight distance of the direct path, and the center frequency of the band of interest, respectively. The calculated power loss is $-80.8$ dB, which deviates from the measured one by $3.1$ dB. This difference is due to frequency-dependent antenna gain values of the \gls{tx} and \gls{rx} antennae.

The assumptions behind measuring small-scale fading using the \gls{cfr} are also validated. The first assumption is that the path coefficients in \eqref{eq:tapped_delay} remains roughly constant with change in frequency. This can be tested indirectly by observing the relation between narrowband power $\rho$ and frequency $f$. The empirical pathloss formula states that $\rho \propto f^{-\eta}$ where $\eta$ is the empirical pathloss coefficient, such that $\eta = 0$ corresponds to no dependence on frequency and $\eta = 2$ is free-space pathloss~\cite{Goldsmith2005}. Fitting the pathloss coefficient to the data shows that $\eta \in [-0.4,1.8]$ for all locations with a median value of $0.7$, indicating some dependency on frequency but less than for free-space propagation. The second assumption is that the range of frequencies, here $8$ GHz, is large compared to the coherence bandwidth. The coherence bandwidth can be approximated by $B_c = 1/D$, where $D$ is the delay spread. Setting the threshold of $-110$ dB for peaks in the \gls{cir} (e.g., as in Fig. \ref{fig:cir_0}) gives that the coherence bandwidth is between $1.6$ MHz and $32.8$ MHz for all locations, indicating that the range of frequencies is sufficient for the narrowband powers to simulate small-scale fading as discussed in Sec. \ref{subsec:small_scale}.

The data is visualized in Fig. \ref{fig:quantile_map} (left) as the estimated $\epsilon$-quantiles in \eqref{eq:quantile_est} of the fading samples at each location.

\subsection{Results}
Following the scenario in Sec. \ref{subsec:problem}, each location in the training data has $N = 8001$ observed fading values, and we select $\epsilon = 1\%$ as the target outage probability with $\delta = 5\%$ as the target meta-probability, i.e., a confidence of $95\%$. Although \gls{urllc} applications typically have error probabilities significantly lower than this, accurately estimating the $\epsilon$-outage capacity for small $\epsilon$ requires a much larger number of samples.
Nevertheless, $\epsilon = 1\%$ allows us to study outage events beyond the average while maintaining accurate estimation of channel statistics.

The $127$ locations are randomly split into a training and a testing dataset, such that $D$ locations are used to generate the \gls{cdi} map, and $127-D$ locations are used for testing. The map models the predictive mean and variance at the test locations according to \eqref{eq:pred_dist}. 
An example of the predictive mean (in dB) for a given realization of the sampling locations is shown in Fig. \ref{fig:quantile_map} (right), where the dots indicate the locations in the training dataset. It can be seen that the \gls{cdi} map generally predicts the quantiles well, except, e.g., in corners of the buildings where the power is significantly lower than at the surrounding locations. For each test location $\mbf{x}_i$ for $i = 1,\dots, 127 - D$, we compute the predicted rate $R(\mbf{x}_i \cond \mathcal{D})$ according to \eqref{eq:rate_select}, and we evaluate the resulting outage probability $p_{\text{out},\mathbf{x}_i}(R(\mbf{x}_i\cond\mathcal{D}))$ and the empirical normalized throughput $\tilde{R}_{\epsilon}(\mbf{x}_i)$ from the test data. This process is repeated $L$ times by re-drawing the $D$ training locations, thus generating a large amount of data to capture the spatial randomness in the scenario. We use $L = 10^4$ throughout the section.

As baseline, we consider the model-based rate selection method in \cite{Angjelichinoski2019statistical} based on $M=10$ fading measurements at the test location $\mbf{x}$. This method ensures that the meta-probability requirement is met under the assumption of Rayleigh fading. Since the number of samples is limited, the scenario reflects, e.g., a mobile scenario where there is limited time and resources to measure the channel statistics. We refer to \cite[Sec. IV-A]{Angjelichinoski2019statistical} for further details on the baseline.

\begin{figure}
\vspace{-5mm}
    \centering
    \includegraphics[scale = 0.5]{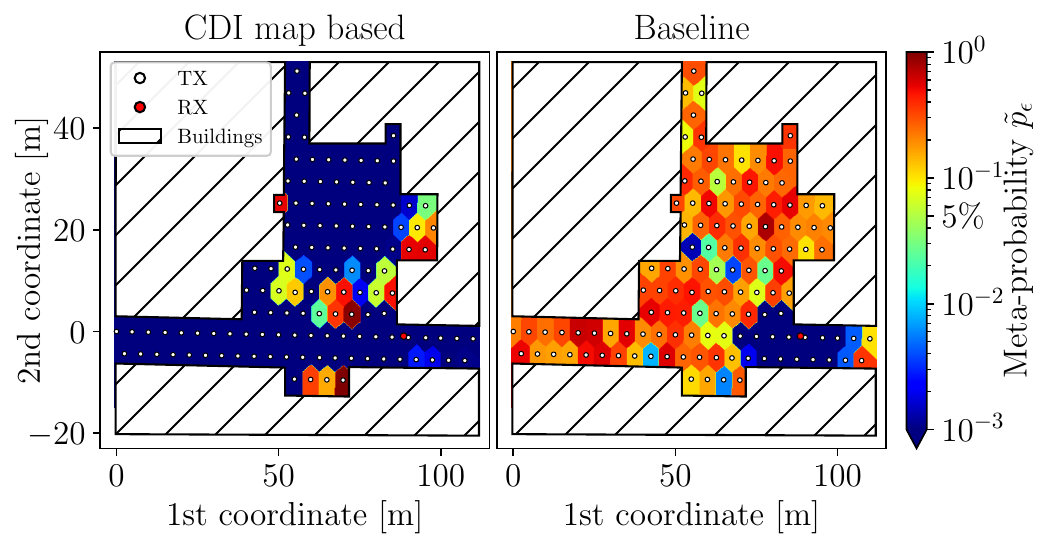}
    \caption{Meta-probability conditioned on \gls{ue} location interpolated based on closest measured point.}
    \label{fig:meta_location}
    \vspace{-3mm}
\end{figure}

We start by analyzing the results spatially by computing the meta-probability conditioned on each location in the data. Figure \ref{fig:meta_location} shows the case of $D = 50$ training locations (averaged across $L$ repetitions).
The \gls{cdi}-based approach generally succeeds in achieving a meta-probability lower than the target of $5\%$, except for certain areas with high meta probabilities. Comparing this to Fig. \ref{fig:quantile_map} (right), it can be seen that a high meta-probability correlates with areas where the $\epsilon$-quantile of the \gls{snr} is significantly lower than its surroundings. For example, the meta-probability is high at the corner location $(50, 25)$ m, where the quantile is more than $4$ dB lower than the surrounding measurements due to the \gls{tx} being placed right next to a concrete wall. This is an example of a scenario in which the Gaussian model underlying the \gls{cdi} map fails to properly predict the quantile, which in turn causes a higher outage probability than desired. On the other hand, the baseline tends to exceed the target meta-probability at most locations, except those close to the \gls{rx}. This is because the locations close to the \gls{rx} have a strong line-of-sight component, which violates the Rayleigh fading assumption and leads to pessimistic rates \cite{Angjelichinoski2019statistical}.

\begin{figure}
\vspace{-3mm}
    \centering
    \includegraphics[scale = 0.5]{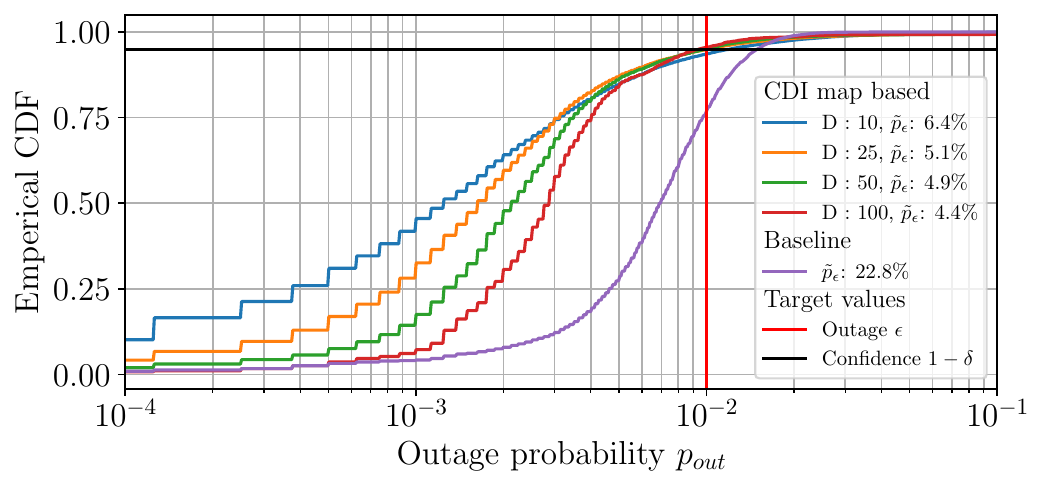}
    \caption{Empirical distribution of outage probabilities. The meta-probabilities are $6.4\%$, $5.1\%$, $4.9\%$ and $4.4\%$ for the \gls{cdi} map based approach with $D = 10, 25, 50$ and $100$, respectively, and $22.8\%$ for the baseline.}
    \label{fig:p_out}
    \vspace{-3mm}
\end{figure}

Next, we evaluate the impact of the number of locations used for training. To this end, we aggregate the resulting outage probabilities across the $127 - D$ test locations and the $L$ repetitions to obtain the distribution of the outage probability across locations. Fig. \ref{fig:p_out} shows the result for $D  = 10, 25, 50$ and  $100$ and the corresponding results obtained using the baseline.
The figure shows that the outage probabilities of the \gls{cdi} map-based approach are generally lower than those of the baseline. The probabilities at $p_{\text{out}} = 1\%$ correspond to $1 - \tilde{p}_{\epsilon}$, and we see that the approach based on the \gls{cdi} map is close to the target meta-probability $\delta = 5\%$ regardless of $D$ with values between $4.4\%$ and $6.4\%$, i.e., a confidence between $93.6\%$ and  $95.6\%$. The figure also shows that increasing $D$ brings the outage probability closer to the target of $1\%$ due to smaller uncertainty of the estimated \gls{cdi} map while still maintaining the desired confidence. This shows that, on average, the \gls{cdi} map is able to predict the channel statistics and also capture the uncertainty of these predictions, thereby allowing it to select a rate that gives the required level of reliability with approximately the desired confidence. The small deviations from $5\%$ occur because the $\epsilon$-quantiles do not exactly follow a Gaussian process and are primarily caused by certain outlier locations, such as the corners of the buildings. 
Although the baseline scheme is closer to the target outage probability, it exceeds the target too frequently with a meta-probability of $22.8\%$. Since the baseline method is designed to meet the specified meta-probability target $\delta$ under Rayleigh fading, we conclude that the Rayleigh assumption falls short of accurately representing the tail of the fading distribution.

\begin{figure}
    \centering
    \includegraphics[scale = 0.5]{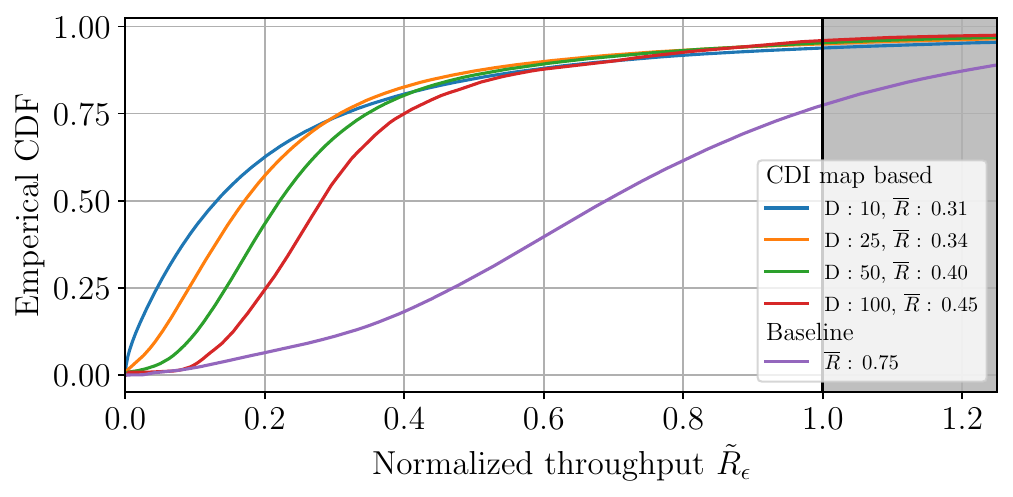}
    \caption{Empirical distribution of normalized throughput. The average normalized throughputs $\overline{R}$ are $0.31$, $0.34$, $0.40$ and $0.45$ for the \gls{cdi} map approach with $D = 10, 25, 50$ and $100$, respectively, and $0.75$ for the baseline.}
    \label{fig:throughput}
    \vspace{-3mm}
\end{figure}

Finally, Fig. \ref{fig:throughput} compares the normalized throughput for a different number of observations $D$ to the normalized throughput obtained by the baseline.
The throughput of the \gls{cdi} map approach is generally lower than that of the baseline, with average values $\overline{R}$ between $0.31$ and $0.45$ compared to $0.75$ for the baseline. However, since a normalized throughput $\tilde{R}_{\epsilon} > 1$ implies that the target reliability has been exceeded, this comes at the cost that the baseline often violates the outage requirement. This highlights a general tradeoff between reliability and throughput. Information about the channel generally helps in selecting the desired rate, and we see that increasing the number of training locations $D$ generally results in a higher average throughput for the \gls{cdi} map approach without increasing the violation frequency. Increasing the number of observations can further decrease the uncertainty, however, the ability to guarantee the target level of reliability even for a limited amount of information ($D = 10$) shows the advantage of the \gls{cdi} map and its suitability for predictive resource allocation in the \gls{urllc} regime.

\section{Conclusion}\label{sec:conclusion}
This paper provided experimental verification of \gls{cdi} maps and their ability to predict rare-event statistics of channel fading by exploiting spatial correlations. A measurement campaign sampled  $127$ locations in a rich scattering environment provided the data for simulating the effect of narrowband small-scale fading. The data was used as input for a location-based rate selection approach, where the communication rate for a \gls{ue} at a previously unobserved location was selected by utilizing an estimated \gls{cdi} map. The results showed that the approach based on \gls{cdi} maps is able to meet the target level of reliability with confidence between $93.6\%$ and $95.6\%$, aiming at $95\%$, whereas the baseline was significantly below the target due to the limited amount of information. There is a tradeoff between reliability and throughput, but increasing the number of measurements used to construct the \gls{cdi} map can increase the throughput while still maintaining high reliability. Achieving a high level of reliability is critical in \gls{urllc} applications, making the \gls{cdi}-based prediction strategy a promising direction in 6G and beyond. Future work on \gls{cdi} maps will include extensions such as the ability to handle non-stationary propagation environments, further exploration of the spatial correlation of \gls{cdi}, and how to model spatio-temporal patterns of interference. 

\section*{Acknowledgment}
This work was supported in part by the Villum Investigator Grant ``WATER'' from the Velux Foundation, Denmark, by
the industrial project ``novel channel sounding techniques for 6G'' founded by Anritsu, and by HORIZON-JU-SNS-2022-STREAM-B-01-03 6G-SHINE project (Grant Agreement No. 101095738). The work of A. E. Kal{\o}r was supported by the Independent Research Fund Denmark under Grant 1056-00006B.

\bibliographystyle{IEEEtran}
\bibliography{references}
\end{document}